\documentclass[11pt,preprint,prd,aps,superscriptaddress,nofootinbib,showpacs,a4paper]{revtex4}

\usepackage[dvips]{graphicx}
\usepackage{amsfonts, color, float, bbm, amsmath}
\usepackage{amssymb}
\usepackage{pstricks}

\newcommand{\braket}[2]{\left\langle#1 |  #2\right\rangle}
\newcommand{\sand}[3]{\langle#1|#2|#3 \rangle}
\def\L{(1-\gamma_5)}

\def\gm {\gamma_\mu}
\def\gM {\gamma^\mu}
\def\gn {\gamma_\nu}
\def\gN {\gamma^\nu}

\def\beq {\begin{equation}}
\def\eeq {\end{equation}}
\def\bea {\begin{eqnarray}}
\def\eea {\end{eqnarray}}

\def\nn {\nonumber}

\def\H {\mathcal{H}_{\mbox{\footnotesize eff}}}
\def\Dppp{D^+\to \pi^+ \pi^- \pi^+}

\def\pp {(\pi^+\pi^-)_S}

\begin{document}
\title{Scalar resonances in a unitary $\pi\pi$ $S$-wave model for $\Dppp$}

\author{D. R. Boito}
\email[Corresponding author:\;]{boito@ifae.es}
\thanks{present address: IFAE, Barcelona, Spain.}
\affiliation{Grup de F\'{\i}sica Te\`orica and IFAE, Universitat
 Aut\`onoma de Barcelona, E-08193 Bellaterra (Barcelona), Spain.}
\affiliation{Instituto de F\'{\i}sica, Universidade de S\~{a}o Paulo,
C.P. 66318, 05315-970, S\~{a}o Paulo, SP, Brazil.}

\author{J.-P. Dedonder}
\affiliation{Laboratoire de Physique Nucl\'eaire et de Hautes \'Energies (IN2P3-CNRS-Universit\'es Paris 6 et 7), Groupe Th\'eorie, Univ. P. \& M. Curie, 4 Pl. Jussieu, F-75252 Paris, France}

\author{B. El-Bennich}
\thanks{Present address: ANL, Argonne, USA.}
\affiliation{Laboratoire de Physique Nucl\'eaire et de Hautes \'Energies (IN2P3-CNRS-Universit\'es Paris 6 et 7), Groupe Th\'eorie, Univ. P. \& M. Curie, 4 Pl. Jussieu, F-75252 Paris, France}
\affiliation{Physics Division, Argonne National Laboratory, Argonne, IL, 60439, USA.}

\author{O. Leitner}
\thanks{Present address: LPNHE, Paris, France.}
\affiliation{Laboratoire de Physique Nucl\'eaire et de Hautes \'Energies (IN2P3-CNRS-Universit\'es Paris 6 et 7), Groupe Th\'eorie, Univ. P. \& M. Curie, 4 Pl. Jussieu, F-75252 Paris, France}
\affiliation{INFN, Laboratori Nazionali di Frascati, Via E. Fermi 40, I-00044 Frascati, Italy}

\author{B. Loiseau}
\affiliation{Laboratoire de Physique Nucl\'eaire et de Hautes \'Energies (IN2P3-CNRS-Universit\'es Paris 6 et 7), Groupe Th\'eorie, Univ. P. \& M. Curie, 4 Pl. Jussieu, F-75252 Paris, France}


\date{\today}

\begin{abstract}
We propose a model for $\Dppp$ decays 
following experimental results which indicate that the two-pion
interaction in the $S$-wave is dominated by the scalar resonances
$f_0(600)/\sigma$ and $f_0(980)$. The weak decay amplitude for $D^+\to
R\, \pi^+$, where $R$ is a resonance that subsequently decays into
$\pi^+\pi^-$, is constructed in a factorization approach. In the
$S$-wave, we implement the strong decay $R\to \pi^+\pi^-$ by means of
a scalar form factor. This provides a unitary description of the
pion-pion interaction in the entire kinematically allowed mass range
$m_{\pi\pi}^2$ from threshold to about 3~GeV$^2$. In order to
reproduce the experimental Dalitz plot for $\Dppp$, we include
contributions beyond the $S$-wave. For the $P$-wave, dominated by the
$\rho(770)^0$, we use a Breit-Wigner description. Higher waves are
accounted for by using the usual isobar prescription for the
$f_2(1270)$ and $\rho(1450)^0$.  The major achievement is a good
reproduction of the experimental $m_{\pi\pi}^2$ distribution, and of
the partial as well as the total $\Dppp$ branching ratios. Our values
are generally smaller than the experimental ones. We discuss this
shortcoming and, as a byproduct, we predict a value for the poorly
known $D\to \sigma$ transition form factor at $q^2=m_\pi^2$.
\end{abstract}

\maketitle

\section{Introduction}
In 2001, the E791 collaboration found a very strong evidence for a
light and broad scalar-isoscalar resonance in $\Dppp$ decays
\cite{E791},  confirming  the existence of the  $f_0(600)$ (also referred
to as the $\sigma$). This pioneering work was soon followed by the
authentication of another elusive scalar, the $K^*_0(800)$ (or $\kappa$), in $D^+ \to
K^- \pi^+ \pi^+$ \cite{E791kappa}.  In the past, several analyses of
$\pi\pi$ scattering data already claimed the presence of the
$\sigma$ in the form of a pole close to threshold and with a
large imaginary part \cite{Eve, OllerOset, Kaminski97}. However, its
manifestation in scattering is subtle and one can consider the E791
experiment as the first solid empirical evidence for this resonance.
The understanding of this pole in $\pi\pi$ scattering
has improved considerably thanks to Chiral Perturbation Theory (ChPT)
\cite{ChPT} and to dispersion relations, namely Roy's equations
\cite{Roy} (for recent works see, for instance, \cite{CGL1,CGL2,KP}). In 2006, the $\sigma$ pole was obtained from
a theoretical  analysis combining these two ingredients and yielding the accurate result $\sqrt{s_\sigma} = (441^{+18}_{-16}
-i\,272^{+9}_{-12.5})$ MeV \cite{CCL}. Furthermore, in the last few years,
experimental evidences for the $\sigma$ in other processes such as
$J/\psi \to \omega \pi^+ \pi^-$ \cite{BESsigma} and, notably, from new
analyses of $\Dppp$ decays \cite{FOCUS,CLEO} have been
published. Thus, at present, one can safely state that the
$\sigma$ is the lightest resonance in the hadronic spectrum.

However, in spite of all theoretical progress, a comprehensive description
of the reaction $\Dppp$ is still to be accomplished. The reasons are
twofold. Firstly, the $c$-quark mass lies in an intermediate range of
energy, between the realm of light quarks ($u$, $d$ and $s$) and that
of heavy quarks ($b$ and $t$).  Although, on the one hand, decays of
light mesons such as the kaon can be treated within the framework of
ChPT and, on the other, decays of the $B$ can be calculated within QCD
Factorization \cite{QCDF}, Heavy Quark Effective Theory \cite{HQEFT}
and Soft Collinear Effective Theory \cite{SCET}, no such  rigorous
framework exists for the $D$. Secondly, one deals with a
three-body final state which renders a full treatment of mesonic final state
interactions (FSI) most involved.
 
In the search of a more sound theoretical framework to treat this type
of reaction, experimentalists usually  fit the Dalitz plot for the
three-body decay with the isobar model. Schematically, the model
consists of a trial amplitude of the form
 \beq
 \mathcal{M} = \alpha_{NR} e^{i \phi_{NR}} + \sum_{i=1}^{n} \alpha_{i}e^{i\phi_i} \mathcal{A}_i,
 \label{Atrial}
 \eeq
 where the first term in the r.h.s. corresponds to a non-resonant
 background and the sum runs over all the $n$ resonances that
 contribute to the decay. In Eq.~(\ref{Atrial}), the parameters $\alpha_i$
 and $\phi_i$ are real constants and the sub-amplitudes
 $\mathcal{A}_i$ (depending on invariant masses) whose main ingredients are relativistic Breit-Wigner
 functions (BW), represent the propagation and decay of each
 resonance.  In the course of the analysis, E791 discovered that the
 usual set of $\pi\pi$ resonances was not sufficient  to yield a good fit
 and, therefore, the addition of a  new BW was required.  Leaving its mass
 and width free to float, an improvement of the least square function, $\chi^2$, was obtained
 with the  parameter values  $m_\sigma=
 (478^{+24}_{-23}\pm 17)$ MeV and $\Gamma_\sigma= (324^{+42}_{-41}\pm
 21)$ MeV.    These values, however, depend strongly
 on the specific  BW chosen to fit the Dalitz plot and, in the literature, one 
 finds various choices   of BW propagators.
  Moreover, this simple description for a very broad
 resonance close to a threshold, like the $\sigma$, has
 many deficiencies  that are discussed,  for instance, in Ref. \cite{Oller}.
 
 On the theoretical  side, a possible way of improving Eq.~(\ref{Atrial}) consists in replacing  the BW functions
  with expressions based on the  knowledge of scattering amplitudes. This procedure respects unitarity and 
  reveals the relation between scattering and production experiments. 
 Using a description  of the $S$-wave two-body FSI  within a  unitarized ChPT  framework \cite{UChPT}, 
   Oller has proposed, in  Ref. \cite{Oller}, a modified  version
  of Eq.~(\ref{Atrial}).
   Albeit 
  successful, this model does not tackle the weak vertex of the reaction and, consequently, the 
 constants $\alpha_i$ and $\phi_i$ in Eq.~(1) remain  fit parameters.

Motivated by the study of CP violation,  more detailed versions of Eq.~(1) were produced in the context of three-body hadronic  $B$ decays, in which both the weak and strong interactions are treated: $B\to \pi\pi\pi$  was considered in Refs. \cite{MeissnerGardner, Gardner2} whereas $B\to K \pi \pi$ and $B\to K \bar K K$ were treated in Refs. \cite{Furman,Bruno, Mouss}.  In these works, the weak decay is evaluated with the help of the effective weak hamiltonian within QCD factorization  whereas the hadronic FSI are taken into account by means of unitary $K\pi$ and $\pi\pi$ form factors constrained by scattering data and ChPT.  These models
confront data very well and are precisely the basis of the approach we follow in the present work.
Our main purpose is to test the description of  $\pi^+\pi^-$ pairs in a relative  $S$-wave state  in $\Dppp$.  

 The decay amplitude for $\Dppp$ is constructed as follows. We
assume that the three-body decay is always mediated by a resonance $R$
as suggested by experimental analyses \cite{E791,CLEO,FOCUS}.   Then,
for the $\pi^+ \pi^-$  pairs in an $S$-wave state, denoted hereafter $(\pi^+\pi^-)_S$, we
factorize the decay amplitude $D^+ \to R \pi ^+$ using the effective
weak hamiltonian within na\"ive factorization. 
Afterwards, the
three-body $(\pi^+\pi^-)_S\,  \pi^+$ final state is constructed from the
intermediate $R\, \pi^+$ state employing the $\pi\pi$ scalar form
factors introduced in Ref. \cite{Furman}.   This
ensures that our description of FSI is unitary and includes the
coupling to the $K\bar K$ state. The form factor is based on the
experimental scattering phase shifts $\delta_{\pi\pi}$ and
$\delta_{K\bar K}$ previously studied in Ref. \cite{Kaminski97}. As usual,
we work within the quasi two-body approximation in which interactions
of the remaining $\pi^+$ with the $(\pi^+\pi^-)_S $  pair are
neglected.  This procedure is repeated for $\pi^+ \pi^-$ in a $P$-wave, which is
well approximated by the $\rho(770)^0$ resonance.
Finally,
two higher mass resonances, the $f_2(1270)$ and the $\rho(1450)^0$, are
included phenomenologically using the isobar model. Our final
amplitude is then fitted to the signal function employed by the E791 collaboration
\cite{E791}; this comparison is carried out following a
scheme presented in Ref.~\cite{Oller}.

The present paper completes the description briefly reported in Ref.~\cite{we} and is organized as follows.  In section \ref{weak}, we
discuss the effective weak hamiltonian and the weak amplitudes
employed in our description for the $S$- and $P$-waves. The
construction of the three-body final state is presented in section
\ref{strong}, results for our fits are displayed in section \ref{fits}
and a summary and discussions   are given in section \ref{conclu}.

\section{Weak amplitudes}
 \label{weak}
 
 Our phenomenological description of weak decays  involving  $S$- or $P$-wave resonances  is based on the effective weak hamiltonian,  $\H$, which is obtained by integrating out the heavy degrees of 
 freedom of the Standard Model (SM) lagrangian.  The hamiltonian is written as an Operator Product Expansion (OPE) and reads
 \beq
 \H^{\Delta C=1} =  \frac{G_F}{\sqrt{2}} \sum_i V_{\tiny\mbox{CKM}} \, C_i(\mu) \hat O_i(\mu)  + \mbox{h.c.}\, ,
\eeq
where $G_F=1.16637(1)\times 10^{-5}$ GeV$^{-2}$ \cite{PDG} is the Fermi decay constant, $V_{\tiny\mbox{CKM}}$ are products of CKM matrix elements, $C_i(\mu)$ are  Wilson coefficients and  $\hat O_i(\mu)$  local operators entering the OPE.   Furthermore, $\mu$ is the renormalization scale which, in our case, is taken to be of order  $m_c$. The operators $\hat O_i(\mu)$ represent the local four-quark
weak interaction in the effective theory while  $C_i(\mu)$ describe the hard or short-distance physics and are calculated perturbatively in the full theory, in this case the SM.  The contributions of all particles with mass $m> \mu =m_c$, 
such as the heavier $b$ and $t$ quarks and the $W$ bosons, are included in~$C_i(\mu)$.

In the present case, we need to consider the Cabibbo suppressed
transition $c\to d u\bar d$. Consequently, the tree level matrix elements are
governed by the coefficient $V_{\tiny\mbox{CKM}}^{\footnotesize\mbox{tree}}= V_{cd}V^*_{ud}\equiv \lambda_d$ which, in
Wolfenstein's parametrization \cite{Wolf}, is of $\mathcal{O}(\lambda)$ where
$\lambda=0.2257$ \cite{PDG}.  In principle, our amplitudes receive
contributions from strong penguins as well but, using the unitarity of
the CKM matrix, one sees that these are governed by $V_{\tiny\mbox{CKM}}^P =
V_{cb}V^*_{ub}$ which is of $\mathcal{O}(\lambda^5)$. Thus, penguin
amplitudes are strongly CKM suppressed and can safely be
neglected. This suppression contrasts with the situation found in
analogous $B$ decays where, since one explores a different sector of the
CKM matrix, penguin operators give rise to sizable contributions
\cite{MeissnerGardner}.
 
 Taking into account only tree operators, the matrix element of $\H$ for the decay $D^+ \to R \pi^+$ can be written as
 \beq
 \sand{R\pi^+}{\H}{D^+} =   \frac{G_F}{\sqrt{2}} \lambda_d \,\sum_{i=1}^{2} C_i(\mu) \sand{R\pi^+}{\hat O_i}{D^+}(\mu),
 \label{Hmatrixel}
 \eeq
 where the  operators $\hat {O}_1$ and $\hat {O}_2$ are given by
 \bea
 \hat O_1&=&   \bar  d \gamma_\nu \L c\,  \,\bar u \gN \L d,\nn \\
 \hat O_2&=& \bar u \gn \L c  \,\, \bar d \gN \L d .
 \eea

To deal with the matrix element $\sand{R\pi^+}{\hat O_i}{D^+}(\mu)$ of
 Eq.~(\ref{Hmatrixel}), we assume that factorization at leading order (in $\Lambda_{QCD}/m_c$ and $\alpha_s$) holds
\bea
\sand{R\pi^+}{\H}{D^+} =   \frac{G_F}{\sqrt{2}} \lambda_d   \Big[a_1(\mu)\sand{\pi^+}{\bar u \gn \L d}{0}\sand{R}{ \bar d \gN \L  c}{D^+}\nn \\+\, a_2(\mu)\sand{R}{\bar d \gn \L d}{0}\sand{\pi^+}{ \bar u \gN \L  c}{D^+}  \Big]. \label{fac}
\eea
In this context,   the new coefficients $a_1(\mu)$ and $a_2(\mu)$ which arise  are  expressed in terms of the Wilson
coefficients $C_1(\mu)$ and $C_2(\mu)$ and of the number of colors $N_c=3$ as   
 \beq
 a_1(\mu) =   C_1(\mu) + \frac{1}{N_c}C_2(\mu), \qquad a_2(\mu) =  C_2(\mu) + \frac{1}{N_c}C_1(\mu).
 \eeq
  We then have  products  of  non perturbative hadronic matrix
 elements
 that assume a decomposition in
 terms of Lorentz invariant form factors. At this point, a remark is
 in order. Since $m_c$ is smaller than $m_b$ by roughly a factor
 three, non-factorizable contributions of order $\Lambda_{QCD}/m_c$
 are larger than in $B$ decays. Hence, the factorization approximation
 may be less reliable for $D$ physics.  Factorization, however, was
 applied successfully to two-body $D$ decays in the seminal papers by
 Bauer, Stech and Wirbel \cite{BSW}. Later, it was used in the 80's
 \cite{DD} and 90's \cite{Bediagaetal} to describe three-body hadronic
 $D$ decays and, more recently, in 2001, Dib and Rosenfeld
 \cite{Rosenfeld} have used the factorization approximation to treat
the $D^+\to \sigma \, \pi^+$ decay  in order to obtain, from the then novel E791
 data, the  $\sigma\pi\pi$ coupling and the form factor for the
 transition $D\to \sigma$.  Finally, the transition $D\to f_0(980)$
 was recently studied employing the same factorization scheme
 \cite{f0}. Thus, in spite of the complications introduced by
 the $c$ mass scale, this approach is a reasonable  starting
 point for a phenomenological analysis. Moreover, it enables the
 description to benefit from the treatment  of mesonic FSI
 developed in the context of $B$ decays\footnote{This is the reason
   why we do not adopt the effective mesonic lagrangian of Refs.
   \cite{Burdman, Wise}.}.
 
 \subsection{\boldmath$D^+\to\sigma\pi^+$ and \boldmath$D^+ \to f_0(980) \pi^+$ decays}
 
 From the available experimental analyses it has become clear that the $\pi\pi$
 $S$-wave  gives the most important contribution to the decay
 $\Dppp$.  The $\sigma$ was found to account for almost $50\%$ of the
 decays whereas the $f_0(980)$ accounts only for about $6\%$ \cite{E791}. This picture remains unchanged  with the more recent analyses
 of FOCUS \cite{FOCUS}  and CLEO \cite{CLEO}  (see Table \ref{tab:ffs}  for the values extracted from these papers). In
 addition, there are indications for a sizable component arising from
 the higher mass scalar resonances $f_0(1370)$ and $f_0(1500)$,
 although the data analyses are not conclusive. The E791 analysis includes only one
 state, the $f_0(1370)$, whereas the CLEO collaboration claims that both are necessary
 to yield a good fit within the isobar model.  In our model,
  the $\sigma$ and the $f_0(980)$ enter explicitly in the amplitude
 while one higher mass state close to 1500 MeV  is accounted for  as a pole in
 the $(\pi^+\pi^-)_S$ form factor, discussed in section \ref{strong}.
  
  The exact nature of scalar mesons, i.e. whether their wave function is dominated by $\bar q q$ or
  $\bar q^2 q^2$ states, or even glueballs, is not yet elucidated. In the case of the $\sigma$, the situation
   is even more obscure, since the issue whether it  is a pre-existing quark state or  dynamically generated
    by the strong $S$-wave $\pi\pi$ interactions  is  still under debate. Therefore,  the inclusion of the $\sigma$ in our 
    amplitude in far from obvious.    On the other hand, we know that the $\sigma$ is an isoscalar state and 
no indications of a strange quark component in its wave function exists.   Thus, we assign to the $\sigma$ the minimal 
quark content compatible with these two facts, namely $(u \bar u +d\bar  d)/\sqrt{2}$.  
Concerning the quark content of the $f_0(980)$, it is well established that it has
a non-negligeable strange component. Higher Fock states, e.g. $\bar q^2 q^2$ and $\bar q^2  q^2 g$, may also be important
 to achieve a comprehensive  description of its wave function\footnote{For a  more complete discussion see Ref. \cite{f0}.}.  In the context of heavy meson decays, however, there are indications that the $\bar q q$ component should be  dominant   \cite{Cheng}.
Here, we  consider  the   $f_0(980)$  as pure $\bar q q$  while allowing for an admixture of $s\bar s$:
 \beq
f_0 = \frac{1}{\sqrt{2}} (u\bar u + d\bar d)\,\sin \theta_{\mbox{\footnotesize{mix}}} +  \, s\bar  s\, \cos  \theta_{\mbox{\footnotesize{mix}}},
\eeq
where $\theta_{\mbox{\footnotesize{mix}}} $ is the mixing angle.
   With these definitions, we can compute the weak amplitude for $D^+\to \sigma\pi^+$ and 
 $D^+\to f_0(980)\pi^+$ decays.  
 
 The matrix elements $\sand{R}{\bar q \gm \L q }{0}$ where $R= \sigma$ or $ f_0(980)$ and $q=u,d$ or $s$
 vanish by $C$ invariance. Hence, the decay amplitude has no $a_2$
 contribution (see Eq.~(\ref{fac})). The annihilation topology is likely to be neglected
 since it contains form factors of light mesons evaluated at high
 momentum ($q^2=m_D^2$) \cite{Rosenfeld}. The weak amplitude is then
 purely proportional to $a_1$. Finally, we note that, in the next
 section, we construct the three-body final state from the
 intermediate $\sigma\pi^+$ and $f_0 \pi^+$ states with the help of the $\pi\pi$
 scalar form factor, thereby properly taking into account the strong
  FSI in this channel.

The relevant form factor for the transition $D\to \sigma$, denoted by $F^{D\to \sigma}_0(q^2)$, is defined as\footnote{Throughout this paper, we employ the decompositions and definitions of Refs.~\cite{MeissnerGardner,Gardner2} for the hadronic form factors, unless otherwise stated.},
 \beq
q^\mu \sand{\sigma(p)}{\bar d \gm \L  c}{D^+(P_D)}= -i (m_D^2 - m_\sigma^2) \tilde F^{D\to \sigma}_0(q^2), 
\eeq
where $q^2=(P_D-p)^2$ and 
$\tilde F_0^{D\to \sigma}(q^2)= F_0^{D\to \sigma}(q^2)/\sqrt{2}$.
Furthermore, with the usual decomposition one has
\beq
\sand{\pi^+(p)}{\bar u \gM \L d}{0} = i f_\pi p^\mu.
\eeq
It  is then straightforward, once factorization has been  applied,  to write the amplitude for $D^+ \to \sigma \pi^+$ with the help of Eq.~(\ref{fac})
\beq
\mathcal{A}(D\to \sigma \pi^+)=  \sand{\sigma \pi^+}{\H}{D^+} =\frac{G_F}{\sqrt{2}} \lambda_d\,  a_1(m_c)\, f_\pi \, (m_D^2 - m_\sigma^2)\, \tilde F_0^{D\rightarrow \sigma}(m^2_\pi) . \label{Asigma}
\eeq
  This expression coincides with the one found in Ref.~\cite{Rosenfeld}  and can 
   easily  be  compared with the amplitude for $B\to \sigma \pi$ from Ref.  \cite{MeissnerGardner} as well. However,   one  should note that in Refs.~\cite{Rosenfeld, MeissnerGardner},
      the normalization   factor $1/\sqrt{2}$  is not included in the $\sigma$ wave function.
  Finally, we have computed the strong penguin contributions 
to Eq.~(\ref{Asigma}) and we have checked explicitly that they can be neglected due to CKM suppression.

 For the decay $D^+ \to f_0(980)\, \pi^+$ we obtain the following result
\beq
\mathcal{A}(D^+ \to f_0 \pi^+)= \frac{G_F}{\sqrt{2}} \, \lambda_d\, a_1(m_c) \,f_\pi \, (m_D^2 - m_{f_0}^2)\, \tilde F_0^{D\rightarrow f_0}(m^2_\pi)  \label{Af0},
\eeq
where we have defined
\beq
\tilde F_0^{D\rightarrow f_0}(m^2_\pi) = \frac{\sin   \theta_{\mbox{\footnotesize{mix}}}}{\sqrt{2}} F_0^{D\to f_0}(m_\pi^2).
\eeq

 \subsection{\boldmath$D^+ \to \rho (770)^0 \; \pi^+$ decay}

The evaluation of the amplitude $\mathcal{A}(D^+ \to \rho(770)^0 \pi^+)$ from Eq.~(\ref{fac}) is quite analogous to the
previous ones except that   the contribution of the colour suppressed tree diagram does not vanish. The quark content in this case
is $\rho^0 = (u\bar u - d\bar d)/\sqrt{2}$ (for simplicity we denote by  $\rho^0$ the $\rho(770)^0$).  In the colour allowed term we need to 
consider the transition $D\to\rho$ which can be  parametrized in terms of form factors using the general $P\to V$ amplitude,  where $P$ and $V$ represent a pseudo-scalar and a vector meson respectively: 
 \beq
 q^\mu \sand{V(p_V)}{{j^V_\mu-j^A_\mu}}{P(p_P)} = - i\,  2 m_V \, (\epsilon^* \cdot q) A_0^{P\to V} (q^2).
 \eeq
  Here  $q=p_P-p_V$, $\epsilon^*$ is the vector meson polarization. Furthermore,  $j^V_\mu$  and $j^A_\mu$ represent bilinear vector and axial-vector quark currents, respectively. In the  colour suppressed topology, we need  the transition $D\to \pi$ obtained from the general $P_1\to P_2$ transition
 \begin{widetext}
\beq
  \sand{P_2(p_P')}{j^V_\mu-j^A_\mu}{P_1(p_P)} =  
    \left[  (p_P + p_P')_\mu - \frac{m_1^2 - m_2^2}{q^2}q_\mu   \right] F_+(q^2) 
    +  \frac{m_1^2 - m_2^2}{q^2} q_\mu F_0(q^2).
\eeq
\end{widetext}
We define the vector decay constant as
\beq
\sand{V}{j^V_\mu}{0} = m_V f_V \epsilon^*_\mu,
\eeq
where we have included the factor $m_V$ to have a decay constant $f_V$ with dimension of energy\footnote{This last definition is not the same as in Refs.  \cite{MeissnerGardner, Gardner2}.}.
With these definitions the result reads
\beq
\mathcal{A}(D^+\to \rho^0 \pi^+) = \,- 2\, (\epsilon^* \cdot q) \, \eta^0,
\label{Arho}
\eeq
where
\beq
\eta^0 =  \frac{G_F}{2} \lambda_d \, m_\rho \left[  a_1(m_c) \, f_\pi A_0^{D\to\rho}(m^2_\pi) + a_2(m_c) \, f_\rho F_1^{D\to\pi}(m^2_\rho)  \right],\label{eta0}
\eeq
 and $q = (P_D - p_\rho) = p_\pi$. The minus sign in Eq.~(\ref{Arho}) arises from the fact that the $d\bar d$ component of the $\rho^0$ is the one that intervenes.  This result has the same structure as
 the one found in  Ref. \cite{MeissnerGardner} for the  $B^-\to \rho^0 \pi^-$ decay.

\section{Hadronic Final State Interactions}
 \label{strong}

 In this section we describe the construction of the three-pion final state from the weak amplitudes
 Eqs.~(\ref{Asigma}), (\ref{Af0}) and (\ref{Arho}).  This is done in order to take into account
 the FSI and the three-body phase space. By FSI, we mean the mesonic interactions in the final
 state after  hadronization. 
 
  It is important to define the kinematics we   employ.  We are considering the generic amplitude $\mathcal M\equiv \mathcal{M}(\Dppp)$ with four-momenta labeled as follows
\beq
D^+(p_D) \to \pi^+(p_1) \, \pi^- (p_-) \, \pi^+ (p_2).
\eeq

 Since in the final state we have two identical $\pi^+$, the amplitude has to be symmetric under the exchange  $p_1 \leftrightarrow p_2$.  What is more,  in order to  work with a Lorentz invariant Daliz plot, it is convenient  to define three
 invariant combinations of momenta
 \beq
 s= (p_1 +p_2)^2, \qquad t= (p_- + p_1)^2, \qquad u= (p_- + p_2)^2. \label{kine}
 \eeq
 They correspond  to the usual Mandelstam variables with
 \beq
 s+ t+u = m_D^2 +3 m^2_{\pi^{\pm}},
 \eeq
 so that only two of them are independent. Resonances occur only in the $t$- and $u$-channels since there are no isospin 2 resonances.
 We denote by $\mathcal{A}_R(u,t)$ the amplitude for a decay mediated by a resonance $R$ in the  $u$-channel.
 The final symmetric result $\mathcal{M}_R(u,t)$ is  hence obtained by summing
\beq
\mathcal{M}_R(u,t) = \mathcal{A}_R(u,t) +\mathcal{A}_R(t,u). \label{sym}
\eeq

 \subsection{\boldmath$S$-wave}

The FSI are taken into account by means
 of the $\pi\pi$ scalar form factor.  This method
 was introduced in Ref.  \cite{MeissnerOller} and  was later applied to
 $B\to \pi\pi  \pi$ decays  in the vicinity of the $\sigma$ pole in Ref.  \cite{MeissnerGardner} using the form
 factors obtained in the context of unitarized ChPT \cite{UChPT}.  In
Refs.  \cite{Furman,Bruno} a similar description was used to describe the $f_0(980)$ in $B\to
 \pi\pi K$ and $B\to K \bar K K$ employing a different set of form
 factors that rely on a previous analysis of $\pi\pi$ and $K\bar K$
 scattering data~\cite{Kaminski97}. These form factors are unitary and
 contain both the $\pi\pi$ and $K\bar K$ channels. Here, we briefly
 summarize the model since all the details can be found in
 Ref.~\cite{MeissnerGardner} and in the appendix of Ref.~\cite{Bruno}.

The two pion scalar form factor, $\Gamma^n(x)$, that is relevant to our work is defined as \cite{MeissnerOller}
\beq
\sand{0}{\bar n n}{\pi(p)\pi(p')} = \sqrt{2} B_0 \Gamma^n(x),
\label{pipiFF}
\eeq
where $\bar n n = (\bar u u + \bar d d)/\sqrt{2}$, $B_0$ is proportional to quark condensate $B_0 = -\sand{0}{\bar q q }{0}/f_\pi^2$ and $x=(p+p')^2$. Since we want
 to describe $\Dppp$ and we have defined in Eq.~(\ref{Asigma}) the amplitude for the transition $D^+ \to \sigma \pi^+$, we need to introduce 
a function, $\Pi_{\sigma \pi \pi}(x)$, that describes the $\sigma$ propagation and decay\footnote{$\Pi_{\sigma\pi\pi}(x)$ corresponds to $\Gamma_{\sigma\pi\pi}(x)$ 
of Ref.  \cite{MeissnerGardner}.}, i.e. the final state interactions. The full amplitude $D^+ \to \sigma \pi^+ \to (\pi^+ \pi^-)_S\, \pi^+$ is given by
\beq
 \mathcal{A}_\sigma(u,t)=  \mathcal{A}(D^+\to \sigma \pi^+) \, \Pi_{\sigma \pi \pi}(u) \nn.\nonumber 
\eeq
\noindent The $\sigma  \to (\pi^+ \pi^-)_S$ decay is described without
resorting to BW expressions.    It can be obtained from the complex conjugate
of Eq.~(\ref{pipiFF}) assuming that, close to the   $\sigma$ pole, this resonance gives the dominant
contribution to $\Gamma^{n}(x)$. We have \cite{MeissnerGardner}
\beq
\Pi_{\sigma \pi \pi}(x) =  \sqrt{\frac{2}{3}}\frac{B_0}{\sand{\sigma}{\bar n n}{0}} \Gamma^{n*}(x) = \chi_\sigma \Gamma^{n*}(x). \label{Gammasigma}
\eeq
The normalization constant 
\beq
\chi_\sigma = \sqrt{\frac{2}{3}}\frac{B_0}{\sand{\sigma}{\bar n n}{0}}
\eeq
 is, in principle, unknown as it depends on the matrix element   $\sand{\sigma}{\bar n n}{0}$. 

To further clarify the meaning of $\Pi_{\sigma\pi\pi}(x)$, it is convenient  to consider its analogue  in a BW framework
\beq
\Pi_{\sigma\pi\pi}^{BW} (x)=\frac{g_{\sigma\pi\pi}}{ m_\sigma^2 -x - im_\sigma\Gamma(x)},
\label{GammaBW}
\eeq
where $g_{\sigma\pi\pi}$ is the coupling between the $\sigma$ and the pions.
We can obtain an estimate for $\chi_\sigma$ comparing expressions  (\ref{Gammasigma}) and (\ref{GammaBW}) at $x=m_\sigma^2$. 
 We obtain
 \beq
 \chi_\sigma |\Gamma^{n*}(m_\sigma^2)  | = \frac{g_{\sigma \pi \pi }}{m_\sigma \Gamma(m_\sigma^2)}.
\label{chiBW}
 \eeq
We take the central values $m_\sigma=478$~MeV and $\Gamma_\sigma= 324$~MeV from the E791
 fit and use $g_{\sigma \pi \pi} =2.52$~GeV~ \cite{gsigma}.  This  yields, with the form factor of Ref.~\cite{Furman},\beq
  \chi_\sigma \approx 29\,\,\,  \mbox{GeV}^{-1}.
  \label{chinum}
  \eeq
 
 From the weak amplitude given in Eq.~(\ref{Asigma}) and from the expression of $\Pi_{\sigma \pi \pi}(x)$ the amplitude for  
$D^+ \to \sigma \pi^+ \to (\pi^+ \pi^-)_S\, \pi^+$ reads
 \bea
 \mathcal{A}_\sigma(u,t)&=& \frac{G_F}{\sqrt{2}} \lambda_d \, a_1(m_c) \, f_\pi \, (m_{D^+}^2 - m_\sigma^2)\, \tilde F_0^{D\rightarrow \sigma}(m^2_\pi) \, \chi_\sigma\, \Gamma^{n*}(u),
 \label{Asigmapipi}
 \eea
where the $t$ dependence is implicit. Similarly for the intermediate resonance $f_0(980)$, one has
 \beq
 \mathcal{A}_{f_0}(u,t)=   \frac{G_F}{\sqrt{2}} \lambda_d \, a_1(m_c) \, f_\pi \, (m_{D^+}^2 - m_{f_0}^2)\,\tilde F_0^{D\rightarrow f_0}(m^2_\pi) \, \chi_{f_0} \, \Gamma^{n*}(u).
 \label{Af0pipi}
 \eeq
 For the weak decay amplitude we make use of Eq.~(\ref{Af0}) and one should note that  the normalization, 
 denoted $\chi_{f_0}$, which by virtue of Eq.~(\ref{chiBW}) is proportional to the coupling $g_{f_0\pi\pi}$, differs from the one found in Eq.~(\ref{Asigmapipi}).
 
 Thus far, we have considered only a small  energy range around the resonance poles. In other works, this prescription for the $S$-wave was always used only within a limited region of the spectrum~\cite{MeissnerGardner,Furman}. We want, however, to describe the whole Dalitz plot for $\Dppp$ where the $\pi\pi$ invariant mass range from
 $2\, m_{\pi^\pm}< \sqrt{s}< (m_{D^+} - m_{\pi^\pm})$, i.e. between $280-1700$ MeV. To
 this aim, an ansatz is required to provide us with an expression for the entire $S$-wave. Since the amplitudes  (\ref{Asigmapipi}) and (\ref{Af0pipi}) are both proportional to $\Gamma^{n*}(u)$, we propose the following
 amplitude for the $S$-wave
 \beq
 \mathcal{A}_{S}(u,t) = \frac{G_F}{\sqrt{2}}  \lambda_d \, a_1(m_c) \, f_\pi \, (m_{D^+}^2 - u) \, \chi_{\mbox{\footnotesize{eff}}}\, \Gamma^{n*}(u).
 \label{ASwave}
 \eeq
 In the last  equation, $\chi_{\mbox{\footnotesize{eff}}}$ is a new normalization constant that encompasses all the form factors and normalizations for the scalar resonances. In addition, we have replaced the terms $(m_{D^+}^2-m_R^2)$ by $(m_{D^+}^2-u)$; this $u$-dependence suppresses the contribution of
 higher mass resonances \cite{Furman}.  It is not easy to obtain a good estimate for $\chi_{\mbox{\footnotesize{eff}}}$ since it receives  contributions from all the scalar-isoscalar states. The following lower bound of $\chi_{\mbox{\footnotesize{eff}}}$ results from Eqs.~(\ref{Asigmapipi}), (\ref{Af0pipi}) and (\ref{ASwave}) 
 \[
 \chi_{\mbox{\footnotesize{eff}}} > \frac{ \sin \theta_{\mbox{\footnotesize{mix}}}}{\sqrt{2}} F_0^{D\to f_0}(m_\pi^2)\, \chi_{f_0}.
 \]
 Using $\tilde F_0^{D\to f_0}(m_\pi^2)=0.215$, average value from the two models of Ref.~\cite{f0}, and  $\chi_{f_0}=28.9$ GeV$^{-1}$~\cite{Bruno} one then gets the estimate 
 \beq
 \chi_{\mbox{\footnotesize{eff}}}> 6.2 \,\,\, \mbox{GeV}^{-1}. \label{chieff}
 \eeq
 
  From Eqs.~(\ref{ASwave})  and (\ref{sym}) we can construct
 our final expression for the  $D^+ \to \pp\, \pi^+$ amplitude
 \beq
 \mathcal{M}_S(u,t) =  \frac{G_F}{\sqrt{2}}  \lambda_d \, a_1(m_c) \, f_\pi\, \chi_{\mbox{\footnotesize{eff}}} \left[ (m_{D^+}^2 - u)\Gamma^{n*}(u) +   (m_{D^+}^2 - t)\Gamma^{n*}(t)   \right].
 \label{MS}
 \eeq
The last expression  has   only one  parameter that can be considered as unknown:  $\chi_{\mbox{\footnotesize{eff}}}$. As far as $\Gamma^{n*}(x)$ is concerned,  with $x = u, t$,  we employ  the form factor  of Ref.  \cite{Furman}. It is unitary and takes into account the coupling to the $K\bar K$ channel\footnote{In a two coupled channel ($\pi\pi$, $K \bar K$) description of the final state interactions for $D^+$ decays, the FSI are incorporated in the form factor via the following unitary equation system (Eq.~(11) of Ref.~\cite{Furman}):
\bea
\Gamma_i^{n*}(x)&=R_i^{n}(x) +&\sum_{j=1}^2\langle k_i \vert R_j^{n}(x)G_j(x)T_{ij}(x) \vert k_j \rangle, \nonumber
\eea
where $\vert k_i \rangle$ and $\vert k_j \rangle$ represent
the wave functions of two mesons in the momentum space and the indices
$i,j=1,2$ refer to the $\pi\pi$ and $K\overline K$ channels, respectively.
The matrix $T$ is the two-body scattering matrix and the functions $G_j(x)$ are the free Green's functions. With this definition, the form factor of Eq. (\ref{GammaKLL}) corresponds to $\Gamma^{n*}_1(x)$.
The driving terms entering in these equations are given by
production functions $R_i^{n}(x)$ representing the meson-meson formation
from $q \bar q$ pairs.
Further details can be found  in Ref.~\cite{Furman}.}.
  This form factor is obtained within an on-shell approximation and can be written explicitly in terms of the $S$-wave scattering phase shifts $\delta_{\pi\pi}(x)$ and $\delta_{K\bar K}(x)$
  as  
 \beq
 \Gamma^{n*}(x) = \frac{1}{2}\Big[ R_{\pi\pi}^{n}(x)\left(1 +\eta(x)e^{2i\delta_{\pi\pi}(x)}     \right) 
- i \; R_{K\bar K}^{n}(x) \sqrt{\frac{k_2}{k_1}} \sqrt{1-\eta^2(x)}e^{i[\delta_{\pi\pi}(x) + \delta_{K\bar K}(x)]}      \Big] ,\label{GammaKLL}   
 \eeq
  where   $\eta(x)$ is the inelasticity for  $\pi\pi$ scattering, $k_1=\sqrt{x/4 -m_\pi^2}$ and $k_2=\sqrt{x/4 -m_{K}^2}$.   For $\delta_{\pi\pi}(x)$, $\delta_{K\bar K}(x)$ and $\eta(x)$, we employ the results   of Ref. \cite{Kaminski97}.    In addition, Eq.~(\ref{GammaKLL})  depends on the production functions $R^n_{\pi\pi}(x)$ and $R^n_{K\bar K}(x)$ introduced in  Ref. \cite{MeissnerOller},  which result from  a matching to the ChPT expansion
of $\Gamma^n(x)$.   In practice, these functions can simply be written  as $R_{i}^n(x) = c_i + d_i \,x$.  The real
coefficients $c_i$ and $d_i$, which depend on  the low energy constants of ChPT,  were determined in  Ref. \cite{MeissnerOller} and updated in  Ref. \cite{Lahde}.  However, the  validity of the production functions beyond  $\sim 1.2$ GeV is not guaranteed.  To circumvent this problem, we introduce in our form factor a cut-off $x_{cut}$
above which we saturate the $R_i^n (x)$, namely
\beq
R_i^n(x) = \left\{
\begin{array}{c}
 c_i +d_i\, x   \,\,\,\,\,\,\  \mbox{for} \,\,\, x<x_{cut}, \\
c_i+d_i\, x_{cut}   \,\,\,  \mbox{for} \,\,\, x>x_{cut}.
\end{array} \right.
\label{Rs}
\eeq
This procedure does not affect the unitarity of the form factor but, of course, it introduces an additional  parameter $x_{cut}$ in the model. Our fits are done for different values of $x_{cut}$ in order to carefully  ascertain the dependence  on this  parameter.

 \subsection{\boldmath$P$-wave}

The construction of  the three-pion final state from the intermediate $\rho^0\pi^+$ state is  similar to that done for the $S$-wave. However, in the case of the $\rho^0$ it is not crucial to employ the vector form factor of the pion to describe the FSI. Since the $\rho^0$ is a relatively narrow
 resonance, far from threshold,  that strongly dominates the corresponding form factor, the BW description gives a good approximation. From the coupling of the $\rho^0$
 to the pair $\pi^+ \pi^-$ defined as $\braket{\pi^+(q_+)\pi^-(q_-)}{\rho^0} =g_\rho (q_- -q_+)$~\cite{MeissnerGardner}  and with the use of  Eq.~(\ref{Arho}) we have
 \beq
 \mathcal{A}_{\rho^0}(u,t) = \eta_0 \, (t-s)\,\Pi_{\rho\pi\pi}(u),
 \eeq
 where $\eta_0$ is given
 in Eq.~(\ref{eta0}) and $s$, $t$ and $u$  are defined in Eq.~(\ref{kine}).
 The factor $(t-s)$ comes from the sum over the polarizations of the $\rho^0$ and the function
 $\Pi_{\rho\pi\pi}(u)$ is defined by
 \beq
 \Pi_{\rho\pi\pi}(u)=\frac{g_\rho}{m_\rho^2 -u - i m_\rho \Gamma_\rho(u) }.
 \eeq
 For the running width $\Gamma_\rho(u)$ we take the usual relativistic prescription
 \beq
 \Gamma_\rho(u) = \frac{m_\rho }{\sqrt{u}}\,  \Gamma_\rho^{\mbox{\tiny tot}}\left( \frac{p(u)}{p(m_\rho^2)}   \right)^3,
 \eeq
 where $p(u)=\sqrt{u/4 -m_\pi^2 }$ and $ \Gamma_\rho^{\mbox{\tiny tot}}$ is the total decay width. The final expression for the amplitude of the decay mediated by the $\rho^0$ is then
 \beq
 \mathcal{M}_{\rho^0}(u,t) = \eta_0 \left[   (t-s)\,\Pi_{\rho\pi\pi}(u) +   (u-s)\,\Pi_{\rho\pi\pi}(t)    \right].
 \label{Mrho}
 \eeq

\section{Results}
\label{fits}

\subsection{Parameter values}

 Since the experimental situation of the  $P$-wave in  $D^+ \to \rho^0\, \pi^+$   is
 less controversial than that for the $S$-wave, this channel can be used to ascertain
 the quality of the model. With  Eq.~(\ref{Mrho}) we can  calculate the branching ratio for the  $D^+\to \rho^0 \, \pi^+$ decay and compare it
  to the experimental average \cite{PDG}
  \[
\mathcal{B}^{\mbox{\footnotesize{PDG}}}(D^+ \to \rho^0 \pi^+, \rho^0 \to \pi^+\pi^-) =(8.2 \pm 1.5) \times 10^{-4}.
\]
  For the numerical input needed we take: $a_1(m_c)= 1.15$~\cite{Buras}, $a_2(m_c) = -0.25$~\cite{Buras}, $f_\pi = 130.4$~MeV~\cite{PDG}, $f_\rho = 0.209$ GeV \cite{Bruno}, $F_1^{D\to \pi} (m_\rho^2)= 0.8$ \cite{F1}, $A_0^{D\to \rho}(m_\pi^2) \approx A_0^{D\to \rho}(0) = 0.75$~\cite{A0}, $g_\rho = 5.8$~\cite{MeissnerGardner}. Using the standard formula for the three-body decay rate \cite{PDG} and taking into account the
 symmetry factor $1/2$ we have, using Eq.~(\ref{Mrho}),
 \beq
\mathcal{B}(D^+ \to \rho^0 \pi^+, \rho^0 \to \pi^+\pi^-) =8.63  \times 10^{-4}.
\label{Brhopi}
 \eeq
 Our branching ratio  is rather sensitive  to the value of the form factor $A_0^{D\to \rho}(m_\pi^2)$ which has  an uncertainty of about $20\%$ at $q^2=0$  \cite{A0}. Therefore, the theoretical error associated with our result is large. However, since the central values for the parameters yield a result in agreement with the experimental average, we  keep these values. Thus, the amplitude $\mathcal{M}_{\rho^0}(u,t)$ serves as a benchmark to the determination of the other parameters of our model.

\subsection{Fits to E791 signal function}

  Since we do not have the real data  at our disposal, the fit procedure
consists in reproducing the E791 signal function and comparing our model to it.  To this aim, we closely follow the method of Ref.~\cite{Oller}. Aitala {\it et. al.} in Ref.~\cite{E791} used in their best fit
the trial amplitude Eq.~(\ref{Atrial}) with six resonances, namely the $\sigma$, the $f_0(980)$ and  $f_0(1370)$ whose quantum numbers are $I^G(J^P)=0^+(0^+)$, the $\rho(770)^0$
 and  $\rho(1450)^0$ with $1^+(1^-)$ and, finally the $f_2(1270)$ with $0^+(2^+)$. Let us remind that 
 the amplitude  has also a  complex constant $\alpha_{NR}e^{i\phi_{NR}}$
 identified  with the non-resonant background. 
With these  ingredients, the signal function can be written
 \beq
 \mathcal{M}^{\mbox{\tiny{E791}}}(u,t) =  \alpha_{NR} e^{i\phi_{NR}} + \sum_{i=1}^6 \alpha_i e^{i\phi_{i}} \, \left[ \mathcal{A}^{\mbox{\tiny{E791}}}_i(u,t) + \mathcal{A}^{\mbox{\tiny{E791}}}_i(t,u)     \right],
 \label{ME791}
 \eeq
where the individual amplitudes $\mathcal{A}^{\mbox{\tiny{E791}}}_i(u,t)$ are modeled as
\beq
\mathcal{A}^{\mbox{\tiny{E791}}}_i(u,t) = F_D^J(u)\times F_i^J(u)\times \Omega_i^J(u,t) \times BW_i(u).
\label{AE791}
\eeq
In the last expression,  $F_D^J(u)$ and $F_i^J(u)$ are Blatt-Weisskopf damping factors that depend on the spin $J$ of the resonance, $\Omega_i^J(u,t)$ are
angular factors and $BW_i(u)$ are the Breit-Wigner propagators.  The full expressions for these functions are rather lengthy and
can be found in the original paper~\cite{E791} or, more detailed, in  Ref. \cite{Oller}.    From Eq.~(\ref{ME791}), and employing the fit results
for $\alpha_i$ and $\phi_i$ \cite{E791} as well as the values for masses and widths used by the E791 collaboration we are able  to reproduce the signal function.  Then, we generate
a Dalitz plot with points separated by $0.05$ GeV and normalize this plot to the number of observed signal events, 1124.  Since the  Dalitz plot  is symmetric under  $u\leftrightarrow t$ it is sufficient to fit only half of the plot.

Our complete amplitude for the $S$-wave is given by Eq.~(\ref{MS}) and the decay mediated
by the $\rho^0$ is described  by Eq.~(\ref{Mrho}).
With only these two contributions, however, it is
not possible to achieve a reasonable
reproduction of the E791 signal function. We have to include the two other resonances that 
give sizable contributions to the fit: the $f_2(1270)$ and the $\rho(1450)$  (denoted for simplicity $f_2$ and $\rho'$ respectively).
 This is done with the help of the isobar model,
using the same expressions as those of \cite{E791}.  We  leave free,  in our fit, 
the corresponding magnitudes $\alpha_{f_2}$ and $\alpha_{\rho'}$ and phases $\phi_{f_2}$ and $\phi_{\rho'}$.
The final amplitude for the decay  $\Dppp$ reads then
\beq
\mathcal{M}(u,t) = \mathcal{M}_S(u,t) +\mathcal{M}_{\rho^0}(u,t)  +\mathcal{M}_{\rho'}(u,t) + \mathcal{M}_{f_2}(u,t),
\label{M}
\eeq
where $\mathcal{M}_S(u,t)$ is given by  Eq.~(\ref{MS}) and  $\mathcal{M}_{\rho^0}(u,t)$ by Eq.~(\ref{Mrho}).  For $\mathcal{M}_{\rho'}(u,t)$ we use
\beq
\mathcal{M}_{\rho'}(u,t) = \alpha_{\rho'}e^{i\phi_{\rho'}} \left[ \mathcal{A}^{\mbox{\tiny{E791}}}_{\rho'}(u,t)  + \mathcal{A}^{\mbox{\tiny{E791}}}_{\rho'}(t,u)\right],
\eeq
where the explicit form of $\mathcal{A}_{\rho'}$ is given in Eq.~(\ref{AE791}).
Analogously, for $\mathcal{M}_{f_2}(u,t)$
we use
\beq
\mathcal{M}_{f_2}(u,t)= \alpha_{f_2}e^{i\phi_{f_2}}\left[ \mathcal{A}^{\mbox{\tiny{E791}}}_{f_2}(u,t)  + \mathcal{A}^{\mbox{\tiny{E791}}}_{f_2}(t,u)\right].
\eeq
We shall refer to this model as ``model A".

The amplitude Eq.~(\ref{M}) receives contributions from five different resonances. These are: the three scalars $\sigma$, $f_0(980)$ and $f_0(1500)$ that appear as poles in our $S$-wave form factor Eq.~(\ref{GammaKLL}) \cite{Furman},  the $P$-wave  resonances $\rho(770)$ and $\rho(1450)$ and the $D$-wave represented by the $f_2(1270)$.
Note that we do not include a non-resonant (NR) amplitude. The necessity for such
 an amplitude is controversial. The  E791 collaboration has shown that the inclusion of the $\sigma$ reduces the   contribution of the NR
 background to less than $10\%$. More recently, the CLEO  collaboration did not find any significant evidence for the NR amplitude and an upper limit of   $3.5\%$ was established \cite{CLEO}.
 In addition, the NR amplitude can  be  energy dependent and the simple complex constant evenly spread over
 the whole phase space may be an unreliable model. Most importantly, the $\pi\pi$ scalar form factor $\Gamma^{n*}(x)$ already includes the NR contributions to $\pi\pi$ scattering which
 could generate a double counting of the background.  
 \begin{table}[!ht]
\begin{center}
\caption{ Fits of  model  A, Eq.~(\ref{M}), to E791 signal function Eq.~(\ref{ME791}). Uncertainties are solely statistical.}\vspace{3mm}
\begin{tabular} {c| c c  c}
\hline
\hline
 & $\sqrt{x_{cut}} = 1.0 $ GeV  & $\sqrt{x_{cut}} = 1.2 $ GeV&$\sqrt{x_{cut}} = 1.4 $ GeV    \\ 
\hline
$\chi_{\mbox{\footnotesize{eff}}}$  [GeV$^{-1}$] &   $6.5 \pm 0.3$  &  6.1 $\pm 0.3 $ &   6.0$\pm$ 0.3\\
$\alpha_{f_2}\times 10^{5}$ &   $(4.7 \pm  0.5) $ &  $(4.2 \pm 0.4) $& $(4.2 \pm 0.4)$  \\
$\phi_{f_2}$ (rd) & $-6.03\pm 0.18$   & $-6.11  \pm 0.20$   & $-6.18 \pm 0.20$\\
$\alpha_\rho'\times 10^{6}$ &  $(2.2\pm 0.8) $  & $(2.6 \pm 0.7) $  & $(3.0 \pm 0.6)$\\
$\phi_{\rho'}$  (rd)&  $-0.57\pm 0.27$   & $-0.30\pm 0.20$   &  $-0.31\pm 0.17$   \\
\hline
$\chi^2$/d.o.f.  & 0.20 & 0.22&  0.22\\
\hline
\hline
\end{tabular}
\label{tab:Par}
\end{center}
\end{table}

\begin{figure}[!ht]
\includegraphics[width=0.4\columnwidth,angle=0]{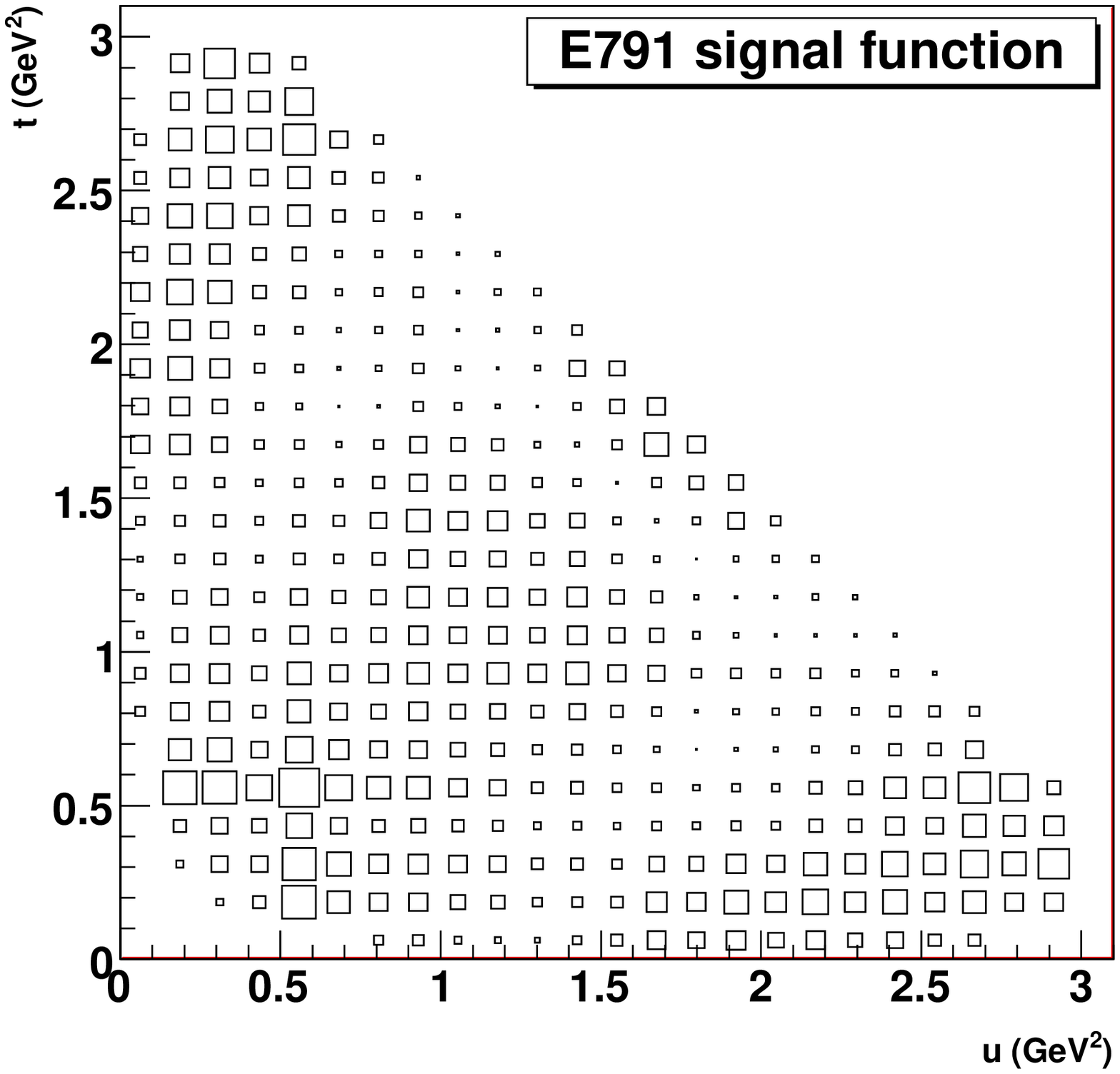}
\includegraphics[width=0.4\columnwidth,angle=0]{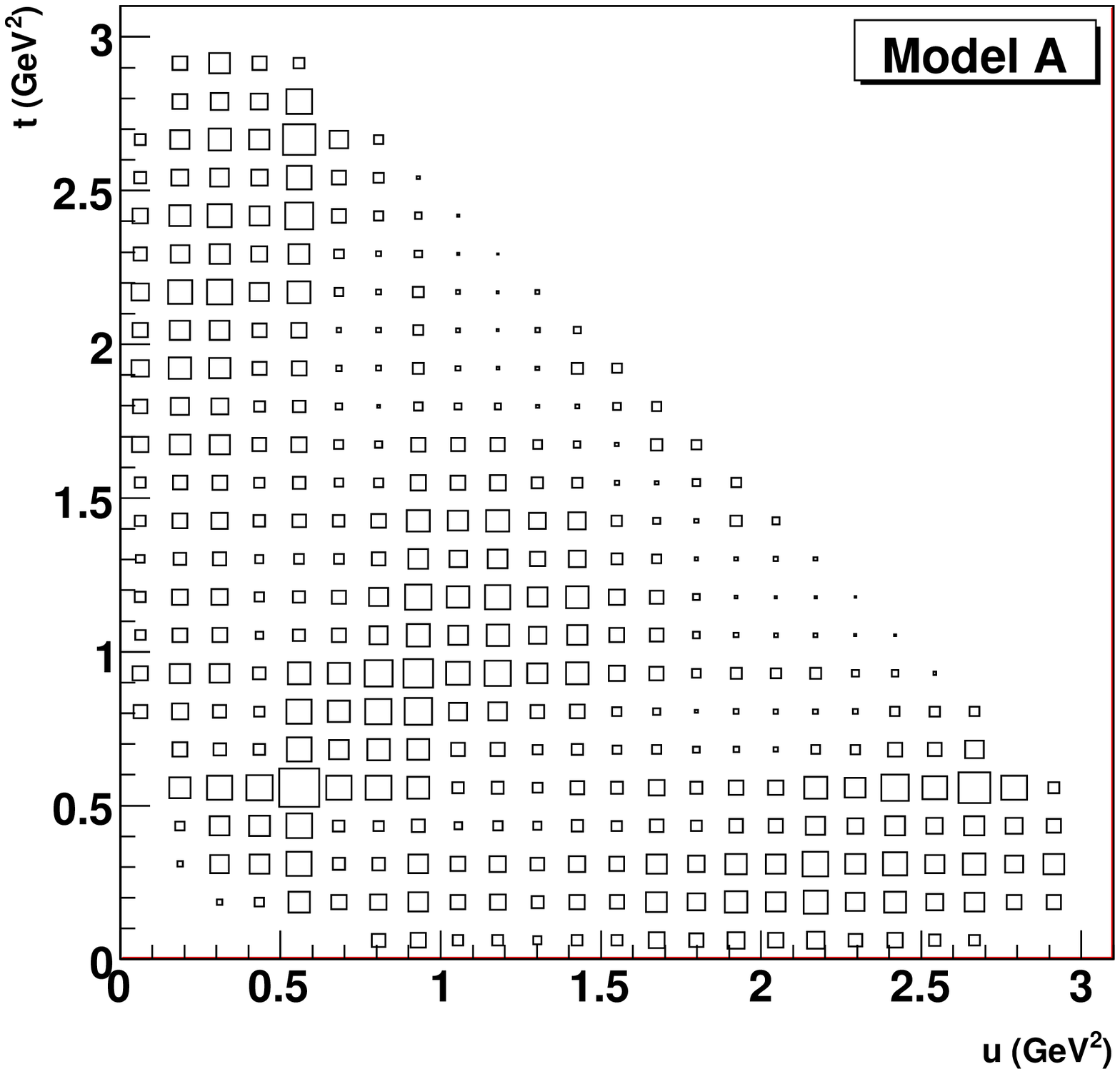}
\caption{{Dalitz plot representation of the signal function employed by E791, Eq.~(\ref{ME791}), (l.h.s),   same
representation for the fitted amplitude given in  Eq.~(\ref{M}) with the parameters of Table \ref{tab:Par} for $\sqrt{x_{cut}}=1.2$ GeV, (r.h.s).} } 
\label{Dalitz}
\end{figure}

\begin{figure}[!ht]
\begin{center}
\includegraphics[width=0.8\columnwidth,angle=0]{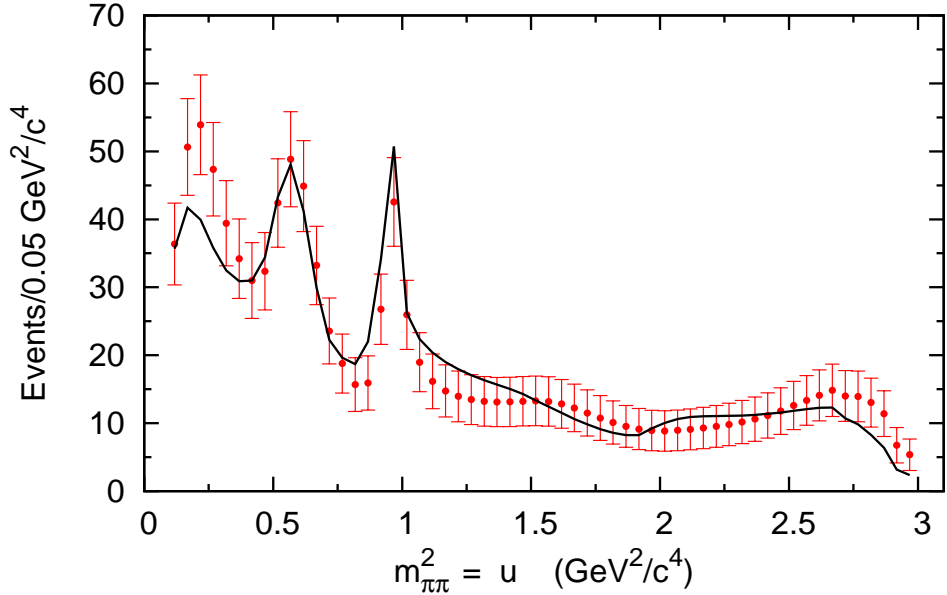}\vspace*{-0.5cm}
\caption{{Projection of the signal function of the E791 collaboration (full dots) and the result of the fit (model A) for the parameters given  in  Table \ref{tab:Par} with $\sqrt{x_{cut}}=1.2$ GeV.} } 
\label{Projec}
\end{center}      
\end{figure}

 Only five free parameters occur in  Eq.~(\ref{M}). Four of them  ($\alpha_{f_2}$, $\phi_{f_2}$, $\alpha_{\rho'}$ and $\phi_{\rho'}$) arise from the
isobar model description of the $f_2(1270)$ and  $\rho(1450)$.  The fifth  is the real constant $\chi_{\mbox{\footnotesize{eff}}}$
 introduced in Eq.~({\ref{ASwave}}). Therefore, the relative weak phase of $S$-wave with respect to the $\rho^0$ is fixed, 
 as well as the  strong phases. 
  In Table \ref{tab:Par} we display the results for fits to the E791 signal function.  
  For the functions $R_{\pi\pi}^n(x)$ and $R^n_{K\bar K}(x)$ of Eq.~(\ref{GammaKLL}) we use  the updated values  obtained in Ref.  \cite{Lahde}.
  We show the results for 
  three fits in which we vary the value of the cut-off $x_{cut}$ introduced in Eq.~(\ref{Rs}). The fit parameters
  do not depend much on $x_{cut}$. When quoting final values, we take the fit with $\sqrt{{x_{cut}}}=1.2$ GeV
  and include an uncertainty due to the dependence on this cut-off.  In Fig. \ref{Dalitz} we show both the E791 signal function and the result of the fit with $\sqrt{x_{cut}}=1.2$ GeV
displayed as  Dalitz plots.  The projection of these two functions is compared in Fig. \ref{Projec}.

Concerning the parameters of the fit, from Table \ref{tab:Par} we see that  the normalization constant  for the $S$-wave, $\chi_{\mbox{\footnotesize{eff}}}$, is 
well determined and confirms our expectations derived in Eq.~(\ref{chieff}). The magnitudes of the $f_2(1270)$ and  of the $\rho(1450)$ are well constrained by the fit as well. On the other hand,  the phases for these higher mass
resonances are not well determined. Finally,   since $\mathcal{M}_{f_2}(u,t)$ and $\mathcal{M}_{\rho'}(u,t)$ in our amplitude Eq.~(\ref{M}) are 
exactly the same as in the function we are fitting to, namely Eq.~(\ref{ME791}), the  interpretation of  the $\chi^2/\mbox{d.o.f}$ as a measurement of the quality of the fit is  not reliable.

From Figs.~\ref{Dalitz} and \ref{Projec}, one  sees that the main discrepancy in the fit comes from the $\sigma$ region. 
This   is most probably due to the off-shell effects that are not included in the form factor given by Eq.~(\ref{GammaKLL}).
The omission of these effects  in $\Pi_{\sigma\pi\pi}(x)$ may lead to an underestimation of the $\sigma$ peak \cite{BR}. The model gives a much better description of the $\rho^0$ and the $f_0(980)$ peaks, both very prominent in the projection. In the higher energy region, where  the $f_2(1270)$
the $\rho(1450)$ and higher mass scalar states are present, our description is reasonable, although not as  good as   in the $\rho^0$ and $f_0(980)$ region.


\subsection{\boldmath$D\to \sigma$  transition form factor}
It is  desirable to extract the form factors for the transition $D^+\to \sigma \pi^+$ and $D^+ \to f_0(980)\pi^+$  from
the fit. The continuous description of the $S$-wave, however, renders this extraction difficult since
the form factors are embedded in   $\chi_{\mbox{\footnotesize{eff}}}$.   Nevertheless, we advance here
a model which aims at determining $F_0^{D\to \sigma}(m_\pi^2)$ and $F_0^{D\to f_0}(m_\pi^2)$.     
We then need to separate the $\sigma$ and $f_0(980)$ contributions in the $S$-wave. The drawback is however  a non-unified  description of $(\pi^+\pi^-)_S$.

 For energies within the elastic domain, $x< 4m_K^2$, the form factor
 of Eq.~(\ref{GammaKLL}) is proportional to $\cos \delta_{\pi\pi}(x)$
 \cite{Furman}. Thus, it has a zero at the point $x_0$ where
 $\delta_{\pi\pi}(x_0)=\pi/2$. Numerically, from the analysis of Ref.
 \cite{Kaminski97}, we have $x_0 \approx (0.828 \mbox{GeV})^2$.  We
 split our $S$-wave amplitude at this point and consider that for
 energies $x<x_0$ the $(\pi^+\pi^-)_S$ is dominated by the transition
 $D\to\sigma$ and hence described by Eq.~(\ref{Asigmapipi}), whereas
 for $x>x_0$ the dominant transition is $D\to f_0(980)$ given by
 Eq.~(\ref{Af0pipi}).  In practice, we substitute Eq.~(\ref{ASwave})
 by Eq.~(\ref{Asigmapipi}) when $x<x_0$ and by Eq.~(\ref{Af0pipi})
 when $x>x_0$. We shall refer to this modified version of the model as
 ``model~B".

 With this modification, we can now consider the products $\chi_\sigma
 \tilde F_0^{D\to \sigma}(m_\pi^2)$ and $\chi_{f_0} \tilde F_0^{D\to
   f_0}(m_\pi^2)$ from Eqs.~(\ref{Asigmapipi}) and (\ref{Af0pipi}) as
 free parameters of the fit. The results of the fit are shown in Table
 \ref{modelB}. The calculated uncertainties take into account the
 three possible sources: statistics, changing the set of $R_i^n(x)$
 and varying $x_{cut}$.  The dominant error in the case of
 $\chi_\sigma \tilde F_0^{D\to \sigma}(m_\pi^2)$ and $\chi_{f_0}
 \tilde F_0^{D\to f_0}(m_\pi^2)$ comes from the two different sets of
 $R_i^n(x)$ functions that we have at our disposal from
 Refs.~\cite{MeissnerOller}~and~\cite{Lahde}. This procedure is conservative,
 since it probably yields an overestimation of the theoretical error.
 The central values are obtained from the most recent determination of
 Ref.~\cite{Lahde}.  In this fit, statistical uncertainties are larger due to the additional parameter of the model indicating that, within the present experimental constraints, the model B is somehow over parametrized as compared to model A. Since the plots for this model are similar to
 Figs. \ref{Dalitz} and \ref{Projec} we refrain from displaying them
 here.

\begin{table}[!ht]
\begin{center}
\caption{ Fit of  model B with $\sqrt{x_{cut}} = 1.2 $ GeV  to E791 signal function Eq.~(\ref{ME791}). Uncertainties include all the possible sources~(see text). }\vspace{3mm}
\begin{tabular} {c|c }
\hline
\hline
$\chi_\sigma\, \tilde F_0^{D\to\sigma}(m_\pi^2)$ &    $\, (8.4\pm 1.4)$ GeV$^{-1}$\\
$\chi_{f_0} \, \tilde F_0^{D\to f_0}(m_\pi^2)$ & $\, (5.6\pm 1.9)$ GeV$^{-1}$ \\
$\alpha_{f_2} \times 10^{5}$ & $5.3\pm0.8$ \\
$\phi_{f_2}$ (rd)  &   $-6.5 \pm 0.4$\\
$\alpha_{\rho'} \times 10^{6}$ & $2.1 \pm 0.9$ \\
$\phi_{\rho'}$  (rd)     &  $-0.2\pm 0.6$\\
\hline
$\chi^2$/d.o.f.  & 0.21\\
\hline
\hline
\end{tabular}
\label{modelB}
\end{center}
\end{table}

Let us now use the results of  Table \ref{modelB} to obtain the transition form factor for $D\to \sigma$. 
In order to disentangle the product $\chi_\sigma \tilde F_0^{D\to \sigma}(m_\pi^2)$ we calculate $\chi_\sigma$ from Eq.~(\ref{chiBW}). Gardner and Mei\ss ner gave 
$\chi_\sigma=20$ GeV$^{-1}$ in Ref. \cite{MeissnerGardner} while employing the production functions $R_i^n(x)$ from Ref.  \cite{Lahde}, we get
 $\chi_\sigma \approx 22$ GeV$^{-1}$.  From Eq.~(\ref{chiBW}) one sees that $\chi_\sigma$ depends on quantities that are not well known such as $\Gamma_{\sigma}$ and $m_{\sigma}$ and we shall  take these values as an indication. Nevertheless, with our value one obtains
\beq
\tilde F_0^{D\to \sigma }(m_\pi^2) = 0.38 \pm0.06.
\eeq 
This result  is to be compared to $0.79\pm 0.15$ from Ref. \cite{Rosenfeld},  $0.57\pm 0.09$ from Ref. \cite{F0sigma1} and $0.42 \pm 0.05$ from Ref. \cite{F0sigma2} 
which is however evaluated at $q^2=0$. Our value is  compatible with the last one, but  smaller than  those 
of Refs. \cite{Rosenfeld, F0sigma1}. More detailed discussions  on the form factor $\tilde F_0^{D\to \sigma}(m_\pi^2)$  are beyond the scope of this work.

Concerning the  $f_0(980)$, we use the recent determination of $F_0^{D\to f_0}(m_\pi^2)$ from the phenomenological analysis of Ref.~\cite{f0} in order to estimate $\chi_{f_0}$. The latter can be compared with the values from Refs.~\cite{Furman, Bruno} to check the consistency of both
approaches.  From Ref.~\cite{f0} we have   
$\tilde F_0^{D\to f_0}(m_\pi^2) = 0.215$ and with the result for the product $\chi_{f_0} \tilde F_0^{D\to f_0 }(m_\pi^2)$  obtained in the fit, this yields 
\beq
\chi_{f_0} = 26 \pm 9\,\,\, \mbox{GeV}^{-1}.
\eeq
This result agrees within uncertainties with the ones employed in $B$ decays: $\chi_{f_0} = 23.5$~GeV$^{-1}$ and   $\chi_{f_0} = 33.5$~GeV$^{-1}$ in the two models of Ref.~\cite{Furman} and $\chi_{f_0} = 28.9$~GeV$^{-1}$  in  Ref. \cite{Bruno}.  This comparison is to be done with care since one always has the product of the normalization $\chi_{f_0}$ and the transition form factors. Therefore, smaller values for $\chi_{f_0}$ can  be compensated  by  larger form factors.

\section{Summary and discussions}
\label{conclu}

It is elucidative  to quantify the weight of the different contributions in our model and compare them  with other results in the literature.
 This is usually  done through the fit fractions, defined, for a given contribution $R$ as,
\beq
f_R =\frac{\int_\mathcal{D} du dt\,   |\mathcal{M}_R(u,t)|^2}{\int_\mathcal{D} du dt\,  |\sum_i \mathcal{M}_i(u,t)|^2},
\eeq
where $\mathcal{D}$ indicates that the integrals are to be performed over the whole Dalitz plot. The fit fractions give an idea of how important a given resonance or partial wave is to the total decay amplitude.  The sum of the  fit fractions 
does not necessarily add up to one due to interference effects. However, one expects that this sum should not deviate widely from unity.  In Table \ref{tab:ffs} we compare the fit fractions of  models A and B
with those of other works as well as with the values quoted by the  PDG \cite{PDG}. In our fit fractions 
we included the uncertainty from $\sqrt{x_{cut}}$  as well as that  arising from the use of  different
sets of $R_i^n(x)$ from Refs.~\cite{MeissnerOller}~and~\cite{Lahde}

\begin{widetext}

\begin{table}[!ht]
\begin{center}
\caption{ Fit fractions (in $\%$) from  E791 \cite{E791} , FOCUS \cite{FOCUS}, CLEO (isobar model) \cite{CLEO},    Oller \cite{Oller}, PDG \cite{PDG} and  models A and B. The uncertainties have been summed quadratically. Results marked with an asterisk are the sum of all $S$-wave contributions and  are 
extrapolated from  the original works. From  Ref. \cite{Oller} we included the $6\%$ of $K\bar K \to (\pi\pi)_S$   in the $(\pi\pi)_S$ fit fraction. }\vspace{3mm}
\begin{tabular} {c| c c c c c c c }
\hline\hline
  & { E791}  & FOCUS& CLEO  & $\,$ Oller $\,$ & {PDG }& {mod. A}  & { mod. B}     \\ 
\hline
$\sigma$      & 46.3$\pm$ 9.2  & - &    41.8$\pm$ 2.9  &      - &  $42.2 \pm 2.7 $  & -  &-\\
NR            & 7.8 $\pm$ 7.8  &   - &     $<$3.5 &   17 &  $<$ 3.5  & - &-\\
$f_0(980)$    & 6.2$\pm$ 1.4  & -&4.1$\pm$ 0.9    &      - &  $4.8 \pm 1.0 $& - &- \\
$f_0(1370)$   & 2.3 $\pm$1.7 &  - &2.6 $\pm$1.9   &   3    &  $ 2.4 \pm 1.3 $& - &-\\
$f_0(1500)$   & - &   -& 3.4 $\pm$1.3  &   -    & $ 3.4 \pm 1.3 $&- &-\\
$\pp$  & $(63\pm 12)^*$ &   56.0 $\pm$ 3.9 &$(51.9\pm 3.8)^*$  &  108 &  $56 \pm 4 $& $50.4 \pm 2.7$  & $56.6\pm 5.2 $\\ 
$\rho(770)$   & 33.6$\pm$3.9  &  30.8$\pm$ 3.9 &20.0$\pm$2.5    &   36  & $25 \pm 4 $& $40\pm  4$  & $34.4 \pm 0.4$\\
$f_2(1270)$    & 19.4$\pm$2.5  &  $11.7 \pm 1.9 $& 18.2$\pm$2.7    &   21 & 15.4 $\pm$ 2.5& 8.6$\pm 2.1$ & $11.8 \pm 0.8$\\
$\rho(1450)$  & 0.7$\pm$0.8  &   -&  $<$2.4  &    1   & $<$ 2.4 & 1.5$\pm 0.6$ & $ 0.8 \pm 0.4$\\
\hline
$\sum_i f_i$ &   116.3                       &  98.5    &        90.1             & 186      &  -         &     100.2   & 103.6    \\
\hline
\hline
\end{tabular}
\label{tab:ffs}
\end{center}
\end{table}

\end{widetext}

To  clarify the content of Table \ref{tab:ffs}, one should remark that the E791 collaboration employed the isobar model \cite{E791}, as already commented. Note   that from the CLEO analyses \cite{CLEO}, we  only quote the isobar model results. FOCUS has performed an analysis where 
 the $S$-wave is described 
continuously  by means  of a $K$-matrix previously determined from $\pi\pi$ scattering data \cite{FOCUS}. This model, therefore, is the closest to ours since both are based on $\pi\pi$ scattering analyses and share the on-shell approximation. Oller's results \cite{Oller}, based on a description of FSI obtained in the context of unitarized ChPT \cite{UChPT}, also rely on previous analysis of $\pi\pi$ scattering and include off-shell effects.
In his model, the non resonant amplitude is kept and, although the resemblance to E791 data is striking,  the price to pay
is a huge interference that makes the sum of fit fractions to be 186\% with 108\% of $\pi\pi$ $S$-wave in the final state. Finally,
we do not sum the fit fractions from the PDG \cite{PDG} since they come from many different analyses. 
Our results are in general agreement with the others. The $S$-wave is within the experimental results and the sum of the fit fractions of our models is very reasonable. 

The CLEO collaboration has performed an analysis where two models based on  the $\pi\pi$ scattering $T$-matrix were applied~\cite{CLEO}. The model of Schechter (non isobar model, 
and  chiral lagrangian with 7 parameters) has similar fit fractions to our results although the coupling to the $K\bar K$ channel is not included in the model. The model
of Achasov (isobar model with 12 parameters) produces results closest to \cite{Oller} with large interference effects. The $(\pi^+\pi^-)_S$ has a fit 
fraction of about $70\%$ an the total sum is roughly $140\%$.

Since we have performed an analysis that starts from the weak vertex, we can calculate the total branching ratio for the decay $\Dppp$ using the fit results for the parameters of the models. This gives
\beq
\begin{array}{c}
\mbox{model A}: \,\,\, \mathcal{B}(\Dppp) =  (2.2^{+0.7}_{-0.5}) \times 10^{-3}, \\
 \mbox{model B}: \,\,\, \mathcal{B}(\Dppp) =  (2.5^{+0.4}_{-0.3}) \times 10^{-3},
 \end{array}
\eeq
whose central values are smaller than the experimental average  \cite{PDG}
\beq
\mathcal{B}^{\mbox{\footnotesize PDG}}(\Dppp) =  (3.21 \pm 0.19) \times 10^{-3}.
\eeq
The results are, however,  coherent with the fact that we miss a part of the contribution in the $\sigma$ region.   Concerning this discrepancy, one can advance some hypothesis. As already stated, this is likely due to the off-shell effects that are not considered in our form factor. Another possible cause for the less prominent   $\sigma$ peak and, consequently, for the smaller branching ratio, are three-body final state interactions.  There are indications of three-body effects in  $D^+ \to K^- \pi^+ \pi^-$ coming from a new analysis technique where the $S$-wave is treated bin by bin in an almost model independent way \cite{Meadows, CLEODKpipi}. The $K\pi$ $S$-wave phase thus obtained is considerably different from the $K\pi$ scattering results and  this has been interpreted as an indication of genuine three-body effects \cite{Alberto}.    To our knowledge, this type of analysis has never been done for $\Dppp$ but, in principle, 
three-body effects could also be important in this case. 

In conclusion, given the relative simplicity of our model, the small number of parameters and the fixed phases between the $S$-wave and the $\rho^0$, the model is able to provide  a fair description of the experimental data for $\Dppp$ within a unitary approach for the $\pi\pi$ final state interactions. Among 
many possible improvements to this work, one can think of introducing $P$ wave form factors to describe the $\rho(770)^0$ and the $\rho(1450)^0$.


\begin{acknowledgments}
We are grateful to Robert Kami\'nski for providing his code to compute the scattering phases  of \cite{Kaminski97}.
We also  thank A.~Furman, M.~R.~Robilotta and R.~Escribano for discussions as well as  J.~A.~Oller and A.~dos~Reis for email exchanges concerning the fit.  DRB thanks the hospitality of LPNHE.
This work was supported by the Department of  Energy, Office of Nuclear Physics, contract no.~DE-AC02-06CH11357, and by a student fellowship 
of the {\em R\'egion \^Ile-de-France}. We also acknowledge partial funding from FAPESP (Brazilian agency) grant no. 04/11154-0 as well 
as from a FAPESP/CNRS bilateral grant, no. 06/50343-8.   The work by DRB  is supported in part by the {\em Ministerio de Educaci\'on y Ciencia}  under grants FPA2005-02211 (FPI scholarship) and    CICYT-FEDER-FPA2008-01430,
the EU Contract No. MRTN-CT-2006-035482, ``FLAVIAnet" and the Spanish Consolider-Ingenio 2010 Programme CPAN (CSD2007-00042).
\end{acknowledgments}

\end{document}